\documentclass{ceab}   
\usepackage{epsfig}     
\usepackage{subfigure}
\usepackage{wrapfig, comment}
\usepackage{graphicx}   
\usepackage{url}
\usepackage{natbib}     
\usepackage[T1]{fontenc} 
\usepackage[UKenglish]{babel}       
    
\def\runningtitle{Mukherjee et al.: ECRS2024 Proceedings}   
\def\articlenumber{7}
\def\ppages{1--10}

\begin{document}

\title{Searching for sub-TeV IceCube neutrinos correlated to sub-threshold GW events}

\author{Tista Mukherjee$^1$\thanks{tista.mukherjee@kit.edu} \hspace*{0.3cm} for the IceCube collaboration$^*$\thanks{https://authorlist.icecube.wisc.edu/icecube}
\vspace{2mm}\\
\it $^1$Institute for Astroparticle Physics, Karlsruhe Institute of Technology\\ }

\maketitle

\begin{abstract}
The IceCube Neutrino Observatory is active in multi-messenger follow-ups of Gravitational Wave (GW) events. Since the release of the Gravitational Wave Transient Catalogue (GWTC)-2.1 by the LIGO-Virgo-KAGRA (LVK) collaboration, sub-threshold GW candidates have been made publicly available. However, they were not followed up in real-time to search for neutrino counterparts. For a deeper under\-standing of these sub-threshold candidates, archival searches are essential. Finding evidence for a neutrino counterpart will enhance the astrophysical significance of these sub-threshold GW candidates and improve their localisation. 
Additionally, it will aid in better understanding possible thresholds on specific GW parameters, beyond which the sub-threshold GW candidates could also be promising multi-messenger sources. Thus, the search contributes to the ongoing efforts to establish correlations between astrophysical neutrino and GW sources.
Here, we present the current status of this ongoing work with the IceCube neutrino data.

\end{abstract}


\section{Introduction}

The IceCube Neutrino Observatory, located at the geographical South Pole, is a cubic-kilometer ice-Cherenkov detector. It is primarily designed to detect high-energy neutrinos with energies exceeding 100 GeV. With all of its resources, IceCube can probe cataclysmic astrophysical sources across the sky with an impressive >99\% detector uptime.
However, the detector's sensitivity was significantly enhanced for low-energy neutrinos with the installation of DeepCore. This denser infill sub-array lowered the energy threshold for neutrino detection by an order of magnitude. This enables IceCube to detect sub-TeV neutrinos, as reported in \cite{deepcore}. 

Leveraging these capabilities, IceCube actively participates in both real-time and archival follow-ups to gravitational wave (GW) alerts issued by the LIGO-Virgo-KAGRA (LVK) collaboration. The results are reported in \cite{greco-gw, GWfollowupO3}. These alerts, from LVK’s first three observation runs (O1–O3), correspond to GW events detected with high confidence with a False Alarm Rate (FAR) < 2 $\mathrm{yr^{-1}}$, depending on background activity around the GW signal detection time. They usually have a high probability of being associated with compact binary coalescences (CBCs). LVK quantifies this by $p_\mathrm{astro}$, which is expected to be greater than 50\%, i.e., $p_\mathrm{astro} \geq$ 0.5. These events are found in the existing Gravitational Wave Transient Catalogues (GWTC) -1, 2.1, 3, published in \cite{gwtc-1, gwtc-2, gwtc-3}, respectively.

In addition to confident detections, GWTC-2.1 and GWTC-3 contain also archival information about sub-threshold events from O3. For these events, the FAR satisfies a more lenient threshold, which is < 2 $\mathrm{day^{-1}}$, resulting in over 2200 candidates. However, no real-time neutrino follow-ups were conducted for these sub-threshold events. Our archival searches now investigate potential neutrino counterparts, examining temporal and spatial correlations of neutrino candidate events within 1000 seconds of each GW detection. In \cite{ICRC23_proceeding}, we presented sensitivity results for 45 sub-threshold candidates from GWTC-2.1. This study now expands the previous one to include additional sub-threshold events from O3b.


\section{The sub-threshold GW candidates}
In this section, we briefly describe the sub-threshold GW candidates detected during O3, which we consider for our analysis. These candidates were identified through different analysis pipelines developed by the LVK collaboration. Four of them are tailored to detect CBCs such as binary black holes (BBH), binary neutron stars (BNS), and neutron star-black hole (NSBH) binaries. These CBC search pipelines are named GstLAL, MBTA, PyCBC, and PyCBC-high mass. Details are described in \cite{gstlal, mbta, pycbc, pycbc-HM} respectively.

Additionally, sub-threshold candidates were identified using the coherent Wave Burst (cWB) all-sky pipeline during the second half of O3 (O3b). Details of the framework is provided in \cite{cwb_allsky}. Unlike model-dependent CBC searches, the cWB-all-sky search looks for plausible sources of short or long-duration GW emissions from poorly modelled sources, such as core-collapse supernovae (CCSNe), long-duration gamma-ray bursts (GRBs), soft gamma repeaters (SGRs), etc, with minimal assumptions about signal morphology. Thus, the cWB all-sky search pipeline can offer the potential to identify GW signals from unforeseen sources.\par
\begin{figure}
\begin{minipage}{0.7\linewidth}
        \includegraphics[width=\textwidth]{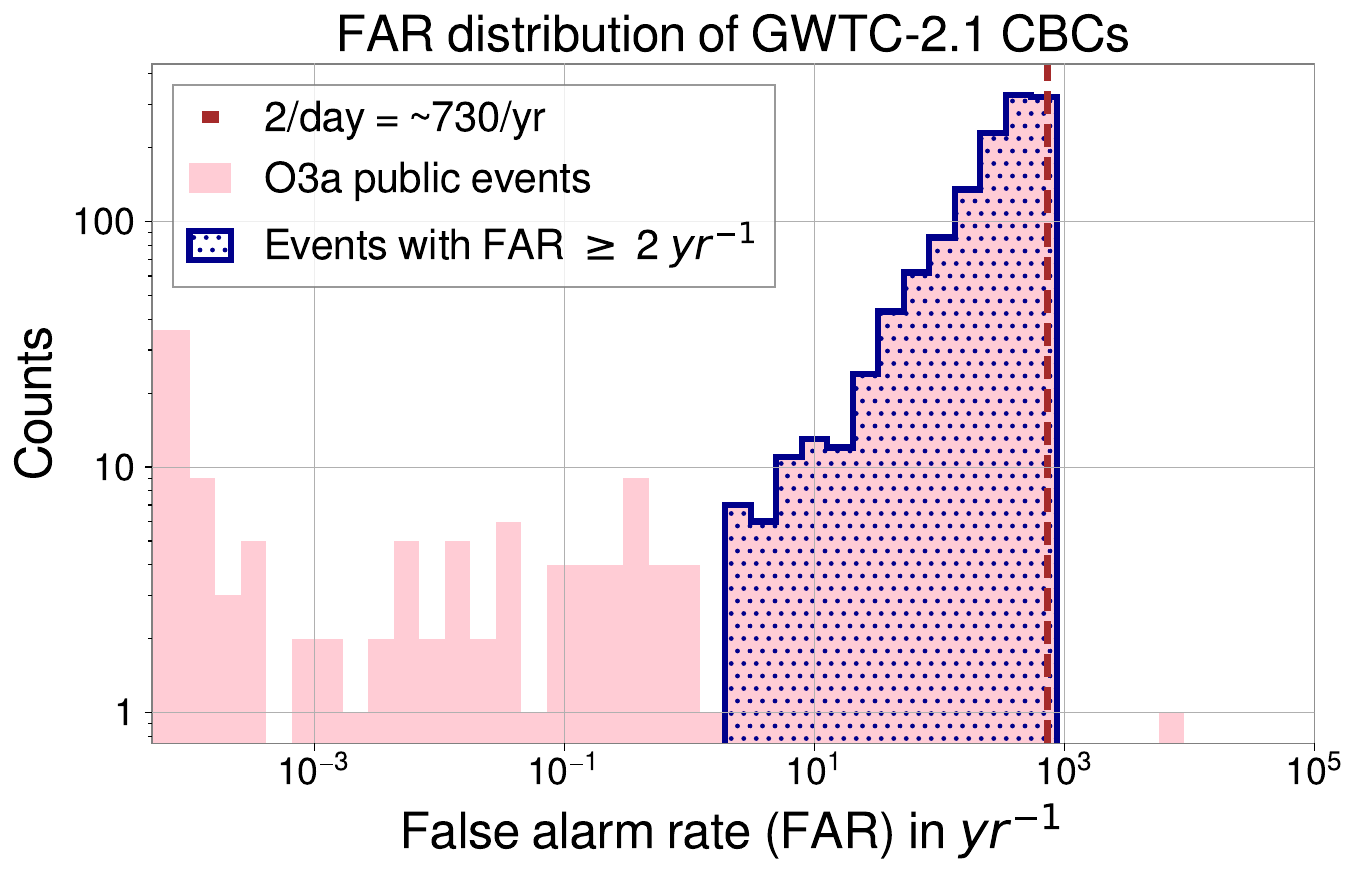}
        \includegraphics[width=\linewidth]{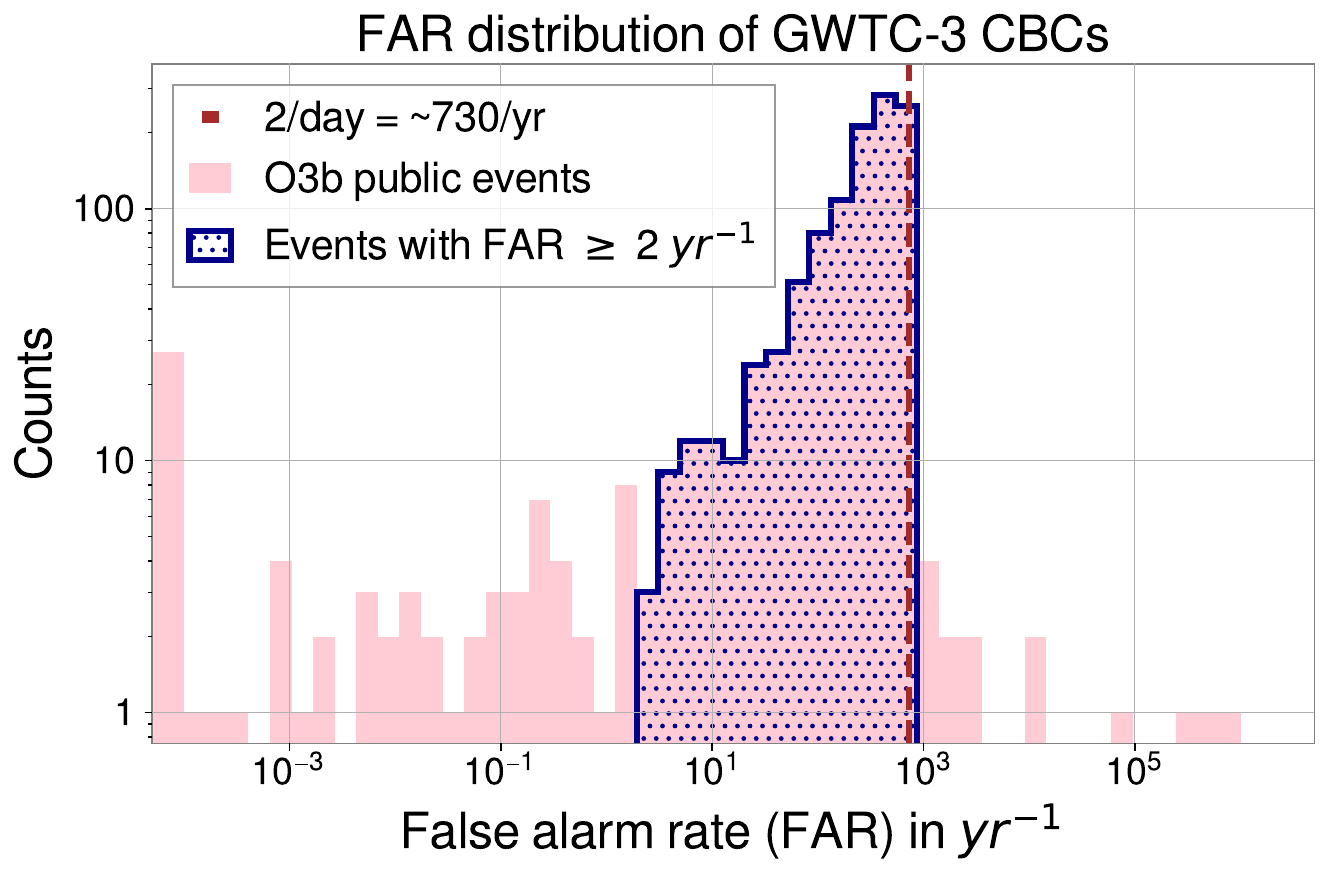}
        \includegraphics[width=\linewidth]{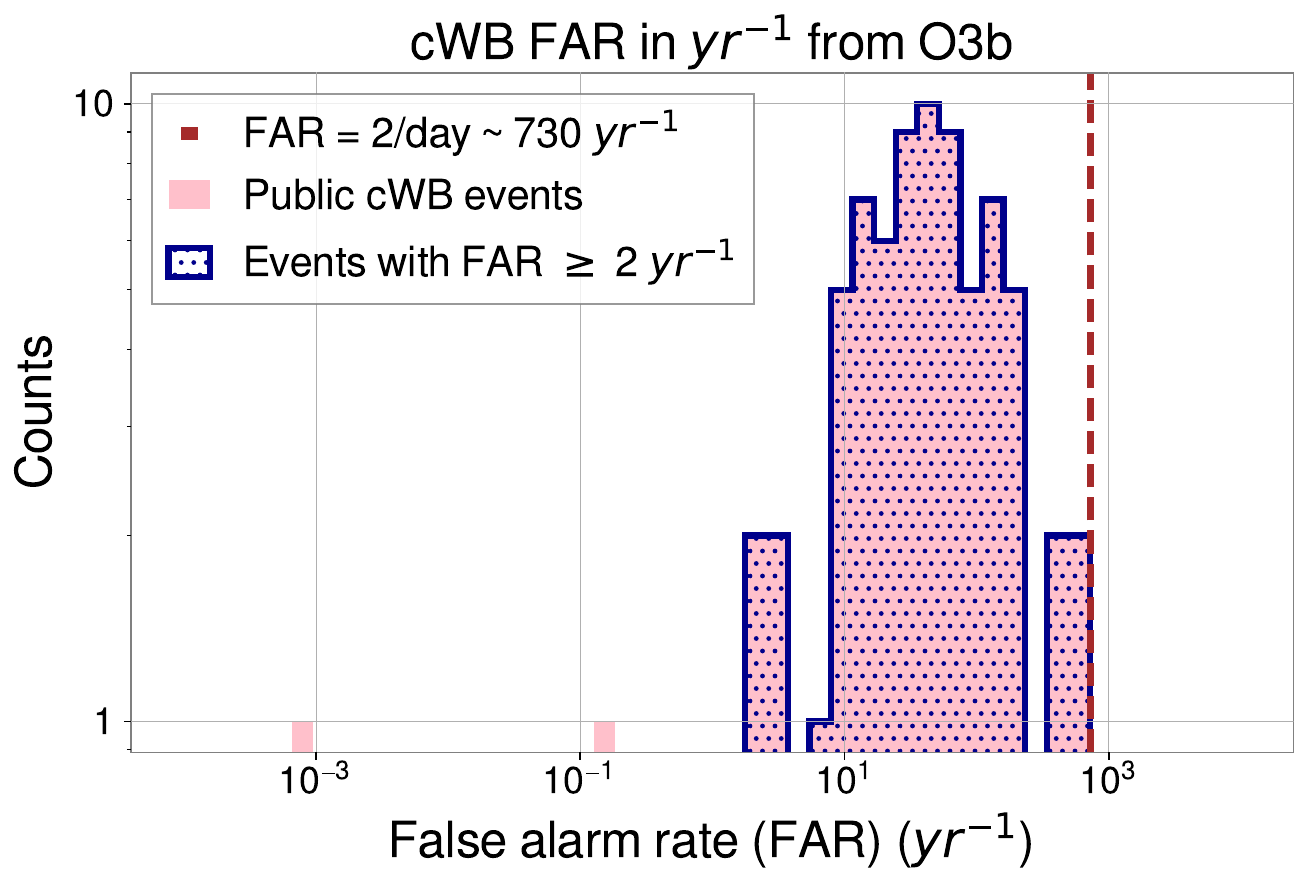}
\end{minipage}
\begin{minipage}{0.29\linewidth}
    \caption{The False Alarm Rate (FAR) distribution for all CBC and cWB candidates from GWTC-2.1 and GWTC-3. These candidates were detected using four template-based and one non-template-based search pipeline during O3. Events with FAR < 2 $\mathrm{yr^{-1}}$ are the confident detections from O3. The dotted region highlights sub-threshold candidates with 2 $\mathrm{yr^{-1}} \leq$ FAR $< 2 ~\mathrm{day^{-1}}$, which are the focus of our analysis. Candidates with FAR values above this range, marked by the dashed line at 2 $\mathrm{day^{-1}}$, are excluded from our catalog search.}
    \label{FAR}
\end{minipage}
\end{figure}

\begin{figure}[!t]
\begin{minipage}{0.49\linewidth}
        \includegraphics[width=\linewidth]{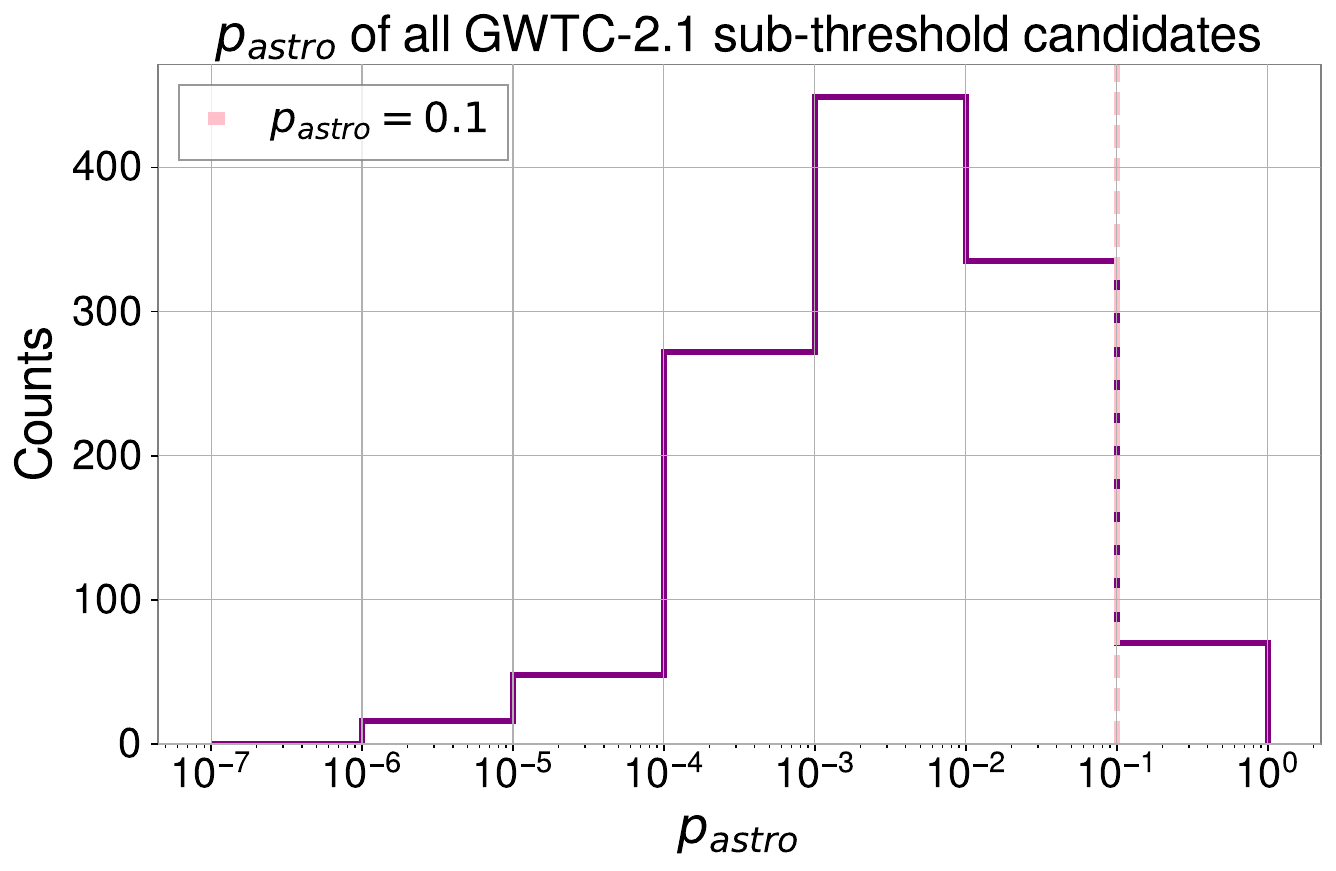}
        \includegraphics[width=\linewidth]{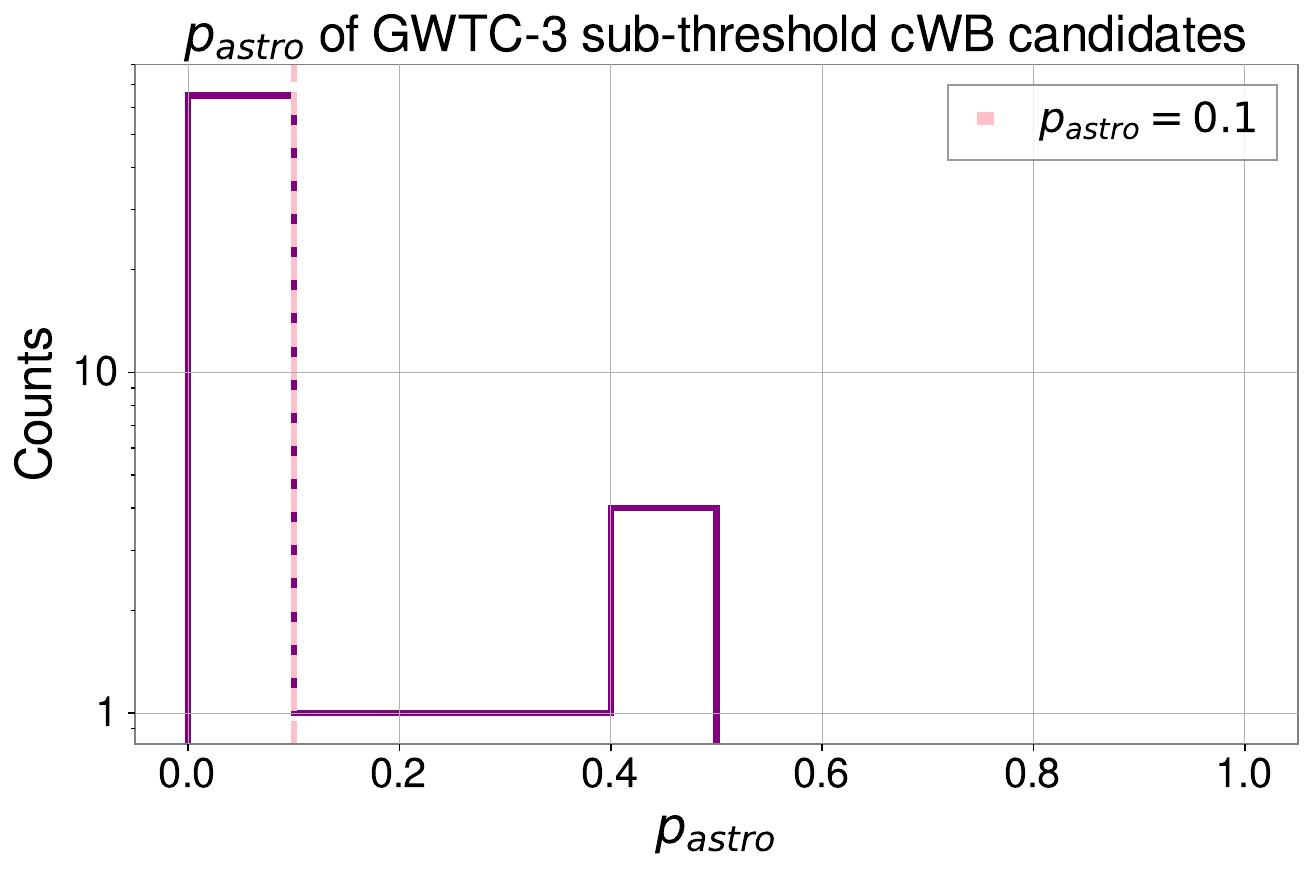}
\end{minipage}
\begin{minipage}{0.49\linewidth}
        \includegraphics[width=\linewidth]{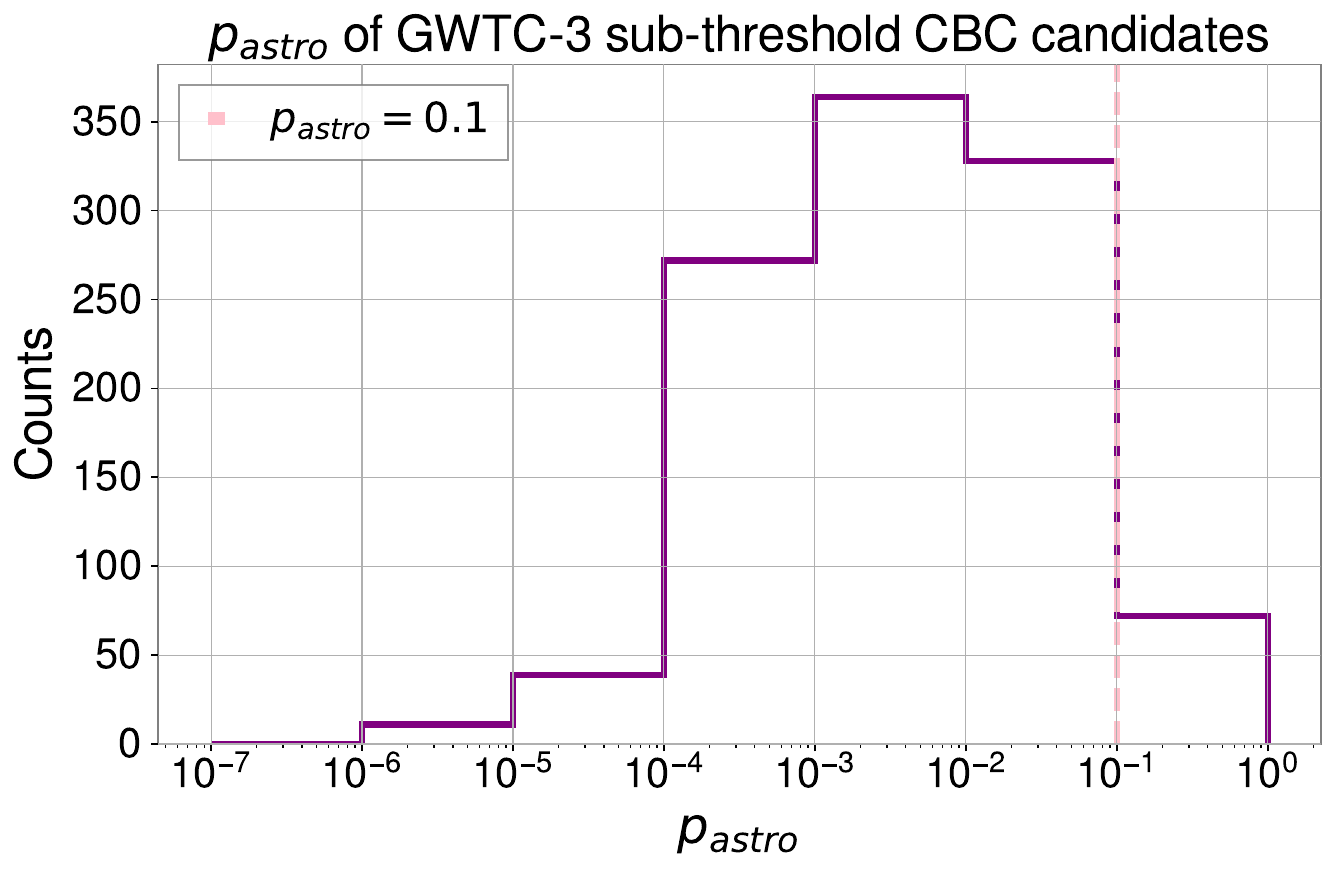}
    \caption{$p_\mathrm{astro}$ distribution for all sub-threshold GW candidates with 2/year $\leq$ FAR < 2/day identified by different CBC and cWB search pipelines during O3a and O3b.
    }
        \label{pastro}
\end{minipage}
\end{figure}

In the GWTC-2.1 and -3, all CBC and cWB candidates with FAR < 2 $\mathrm{day^{-1}}$ (approximately 730 $\mathrm{yr^{-1}}$) were included. Importantly, this means that confident events with FAR < 2 $\mathrm{yr^{-1}}$ are also to be found in this catalogue of sub-threshold candidates. This is evident from Fig. \ref{FAR}. To distinguish the sub-threshold candidates, we first exclude confident GW events that have already been followed up with high- and low-energy IceCube neutrinos, as reported in \cite{greco-gw, GWfollowupO3}.
For the remaining GW candidates with FAR > 2 $\mathrm{yr^{-1}}$, we look into their $p_\mathrm{astro}$ distribution, shown in Fig. \ref{pastro}. More than 90\% of the candidates have $p_\mathrm{astro} \simeq$ 0, precisely $p_\mathrm{astro}$ < 0.1. They will not be included in the catalogue search we are planning to perform.


\begin{figure}[!t]
    \begin{minipage}{0.6\linewidth}
	\includegraphics[width=\linewidth]{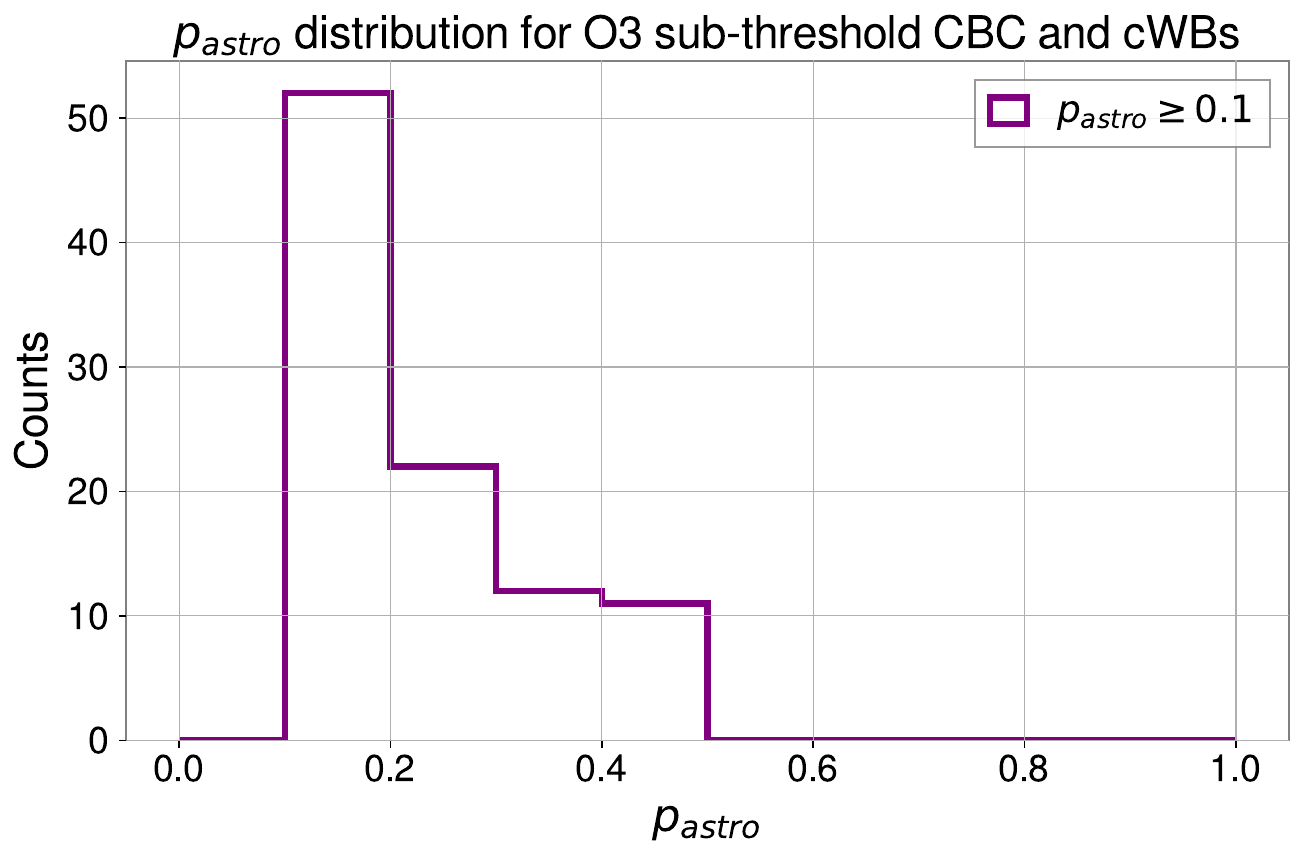}
    \end{minipage}
    \begin{minipage}{0.39\linewidth}
	\caption{$p_\mathrm{astro}$ distribution for the 97 sub-threshold GW candidates, both CBCs and cWBs, detected during O3, to be considered for neutrino counterpart search. These candidates have FAR < 2/day and $p_\mathrm{astro} \geq$ 0.1.}
	\label{pastro_all}
    \end{minipage}
\end{figure}

We further refine our sub-threshold candidate selection by excluding events with $p_\mathrm{astro}$ < 0.1, because they are more likely to be masked by terrestrial noise. Next, we remove any sub-threshold candidates overlapping with confident events already cataloged in GWTC-2.1 and GWTC-3, as these were previously included in IceCube's archival searches. For candidates identified by multiple pipelines, we retain only those with the highest $p_\mathrm{astro}$ from one particular pipeline, at a given event-time. After applying these criteria, we retain 97 unique sub-threshold events: 90 CBCs from O3 and 7 cWB candidates from O3b. The $p_\mathrm{astro}$ distribution for these candidates is shown in Fig. \ref{pastro_all}.

\section{IceCube sub-TeV neutrinos}

The sub-TeV neutrino dataset available within IceCube is called `GRECO' (GeV Reconstructed Events with Containment for Oscillations). This is an all-sky and all-flavour dataset suitable for transient follow-up searches, as introduced in \cite{novae}.
Previous studies (e.g., \cite{greco-gw}) have used GRECO to investigate possible lower-energy neutrino counterparts to 90 confident GW events observed during the O1, O2, and O3 runs. Here, we extend this approach by using GRECO to search for sub-TeV neutrino counterparts associated with selected sub-threshold GW candidates from the O3 run.

\section{Analysis method}

To search for the neutrino counterparts, we will follow an Unbinned Maximum Likelihood (UML) analysis. We define the following likelihood function $\mathcal{L}$ ($n_\mathrm{s}( \gamma$)), which is given as
\begin{equation}
\label{likelihood}
    \mathcal{L} (n_\mathrm{s} (\gamma)) = \frac{(n_\mathrm{s} + n_\mathrm{b})^N}{N!} \mathrm e^{-(n_\mathrm{s} + n_\mathrm{b})} \prod_{i = 1}^{N} \Big ( \frac{n_\mathrm{s} S_\mathrm{i}}{n_\mathrm{s} + n_\mathrm{b}} + \frac{n_\mathrm{b} B_\mathrm{i}}{n_\mathrm{s} + n_\mathrm{b}} \Big ).
\end{equation}

Here, $n_\mathrm{s}$ and $n_\mathrm{b}$ are signal and background neutrino events, respectively, and their sum gives the total number of observed events, $N = n_\mathrm{s} + n_\mathrm{b}$. The index $i$ runs over each of the $N$ candidate neutrino events. The signal probability density function (PDF) for the $i^{th}$ event, denoted as $S_\mathrm{i}$, is constructed using the reconstructed sky position, energy, and angular uncertainty of the event. It also incorporates information about the source position and depends on the spectral index $\gamma$, which characterizes the assumed energy spectrum of the signal neutrinos. The background PDF, $B_\mathrm{i}$, for the $i^{th}$ event is estimated from scrambled data and depends primarily on the event’s declination and reconstructed energy.

Using this likelihood function, we define a test statistic (TS).
\begin{equation}
\label{TS}
    TS = 
    \Bigg[ 2~\mathrm{ln}  \Bigg( \frac{\mathcal{L}_\mathrm{k} (n_\mathrm{s}(\gamma)) \cdot \omega_\mathrm{k}}{\mathcal{L}_\mathrm{k} (n_\mathrm{s} = 0) } \Bigg) \Bigg ].
\end{equation}

Here, $k$ is the index of each pixel in the sky and $\omega_{k}$ is a spatial prior term at that location. For each pixel, $\omega_\mathrm{k}$ is scaled linearly with the probability of having a GW source in that pixel.
The details of each of these components and the entire framework for analysis with sub-threshold GW candidates have been described in \cite{ICRC23_proceeding}.   

\section{Background and sensitivity studies}
As a first step, we construct the background TS (BGTS) distribution for each sub-threshold GW candidate. This is achieved by randomly sampling the arrival times of each neutrino event detected within a $\pm$ 5-day time window around each GW detection time. This procedure is repeated 10,000 times, individually considered as an independent trial. The resulting BGTS distribution is then used as a reference for a background-like case. Then we assess the sensitivity by "injecting" signal-like neutrino events, randomly selected from a Monte Carlo dataset with an assumed spectral index of $\gamma = 2$, onto the scrambled (background-like) dataset. From these injections, we calculate the 90\% sensitivity, which is reported for each selected sub-threshold GW candidate. This sensitivity estimate provides a benchmark for the detection capabilities of our analysis.

\subsection{Background TS distribution}

\begin{figure}[h]
    \centering
    \begin{minipage}{0.5\textwidth}
        \centering
        \includegraphics[width=\textwidth]{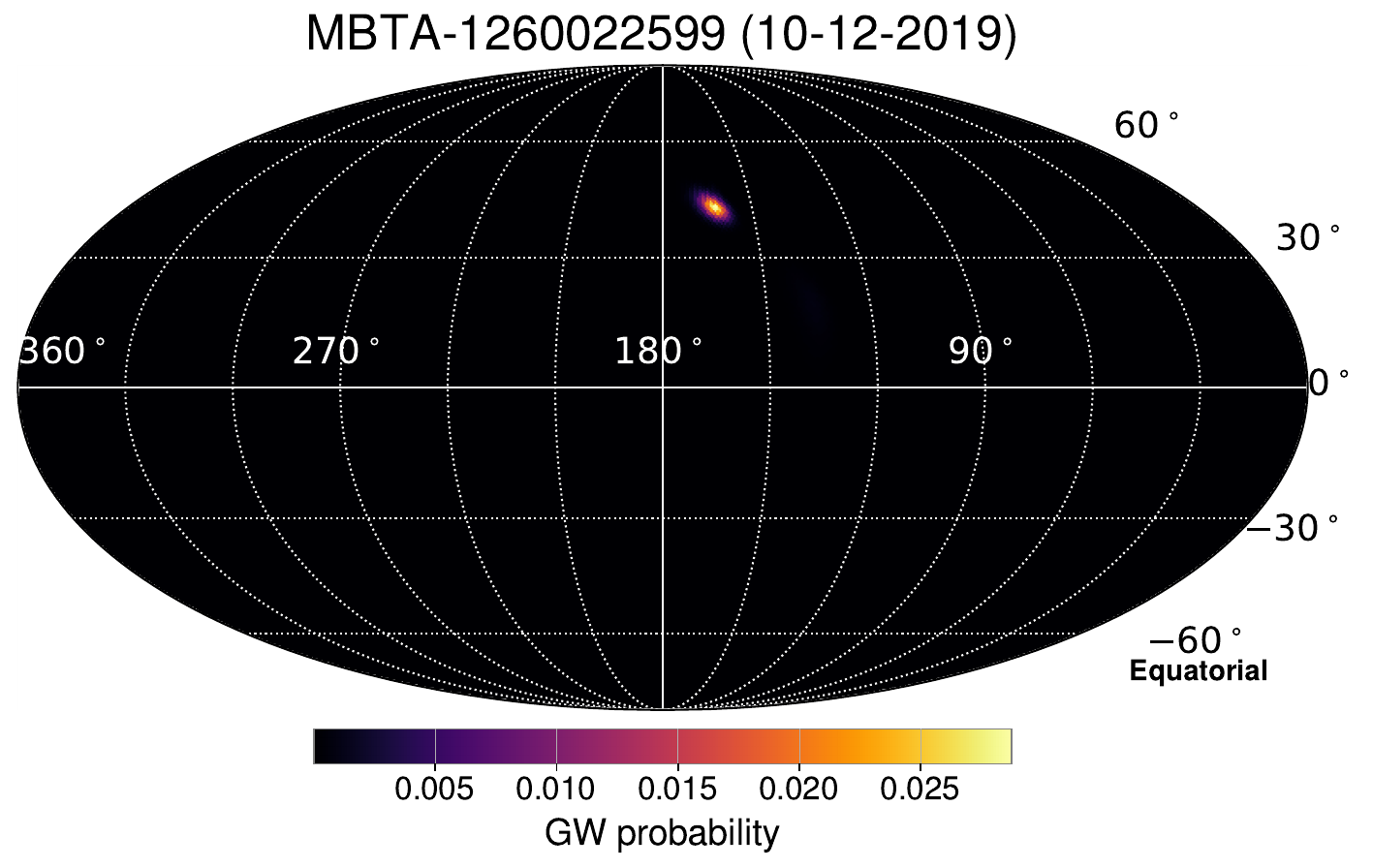}
        \end{minipage}%
    \begin{minipage}{0.5\textwidth}
        \centering
        \includegraphics[width=\textwidth]{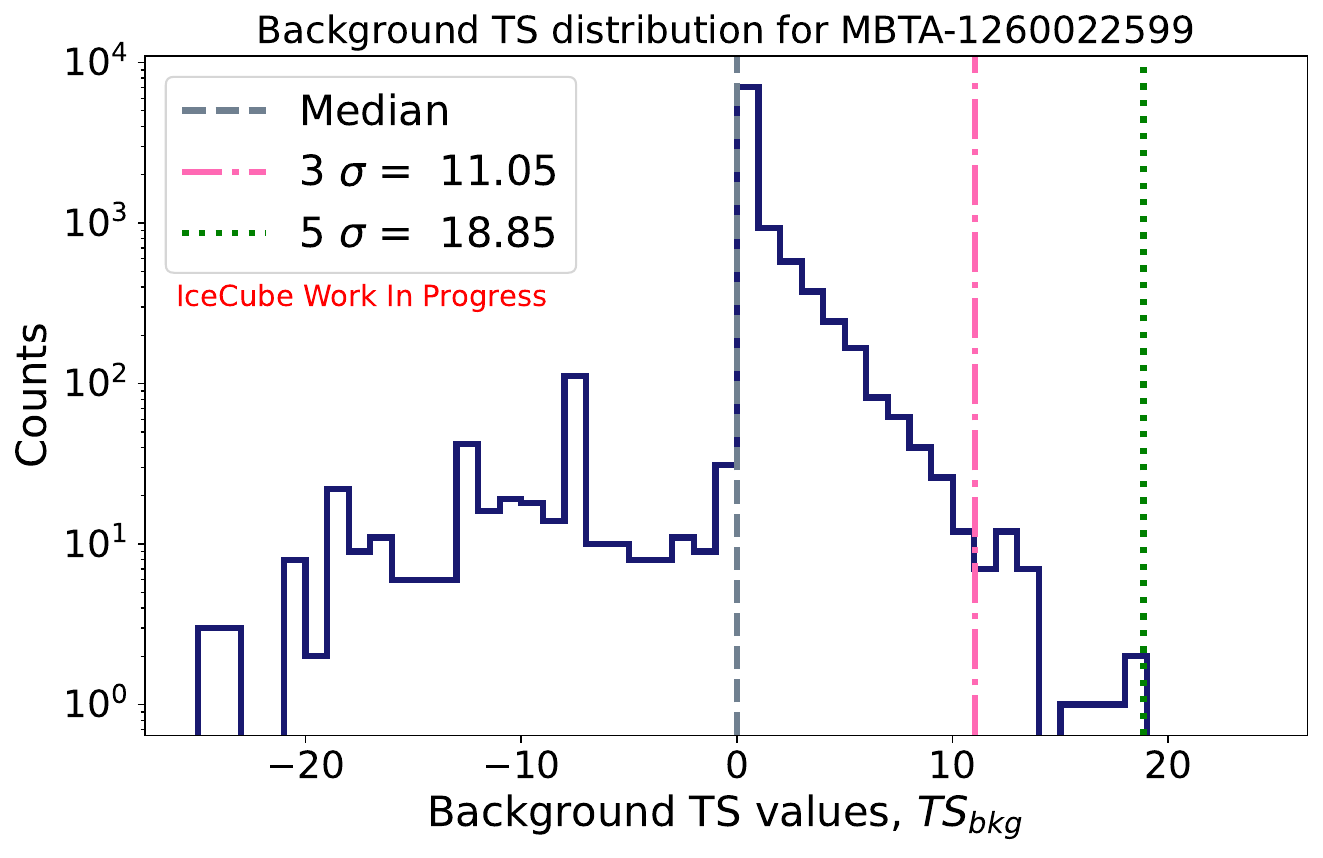}
    \end{minipage}
	\caption{The sub-threshold GW candidate MBTA-1260022599 ($p_\mathrm{astro}$ = 0.4), detected during O3b: (left) the corresponding skymap for this candidate, and (right) the background TS distribution including spatial priors.
 }
        \label{CBC_BkgTS}
\end{figure}

\begin{figure}[htbp]
    \centering
    \begin{minipage}{0.5\textwidth}
        \centering
        \includegraphics[width=\textwidth]{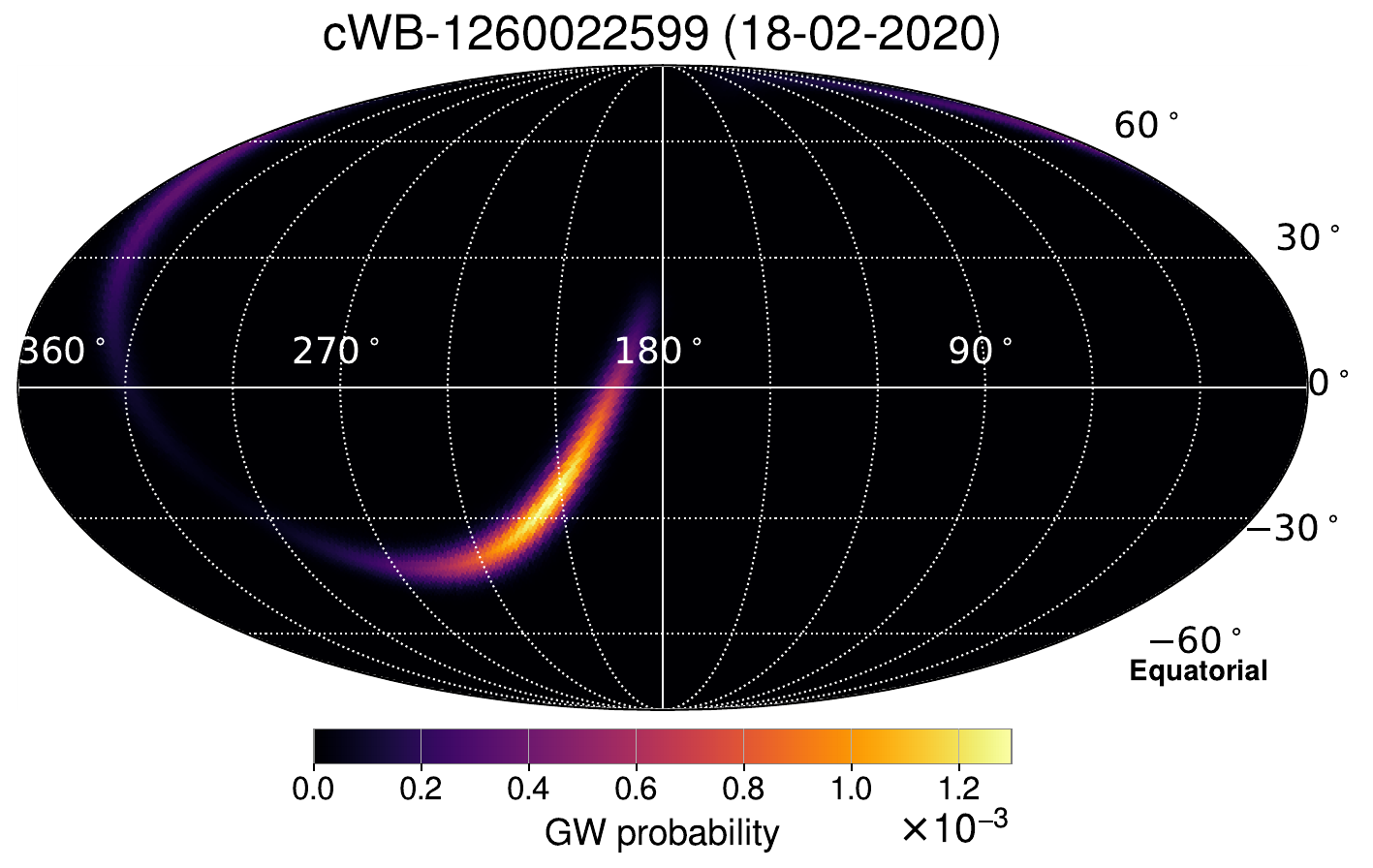}
        \end{minipage}%
    \begin{minipage}{0.5\textwidth}
        \centering
        \includegraphics[width=\textwidth]{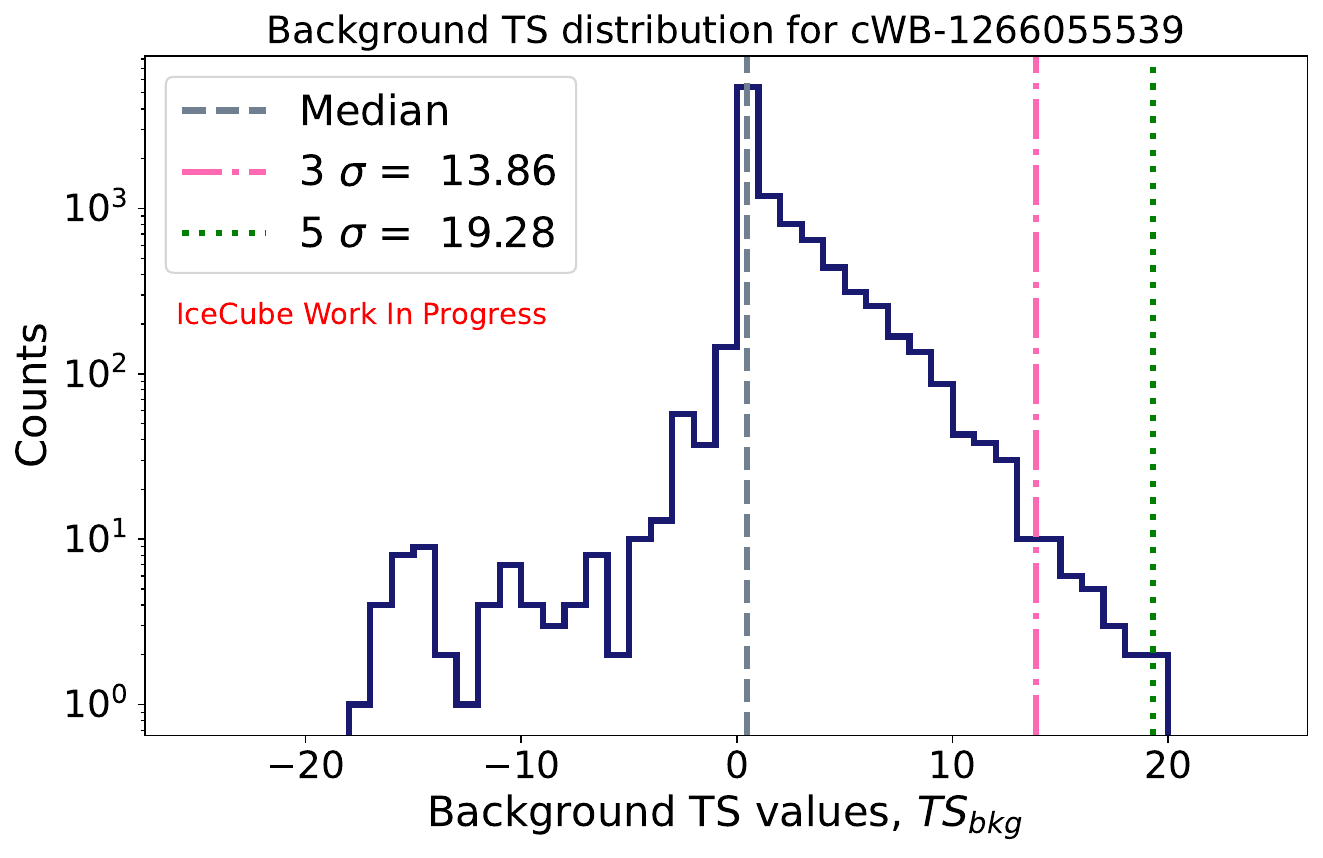}
    \end{minipage}%
	\caption{The sub-threshold GW candidate cWB-1266055539 ($p_\mathrm{astro}$ = 0.49), detected during O3b: (left) the corresponding skymap for this candidate, and (right) the background TS distribution including spatial priors.}
        \label{cWB_BkgTS}
\end{figure}

In Fig. \ref{CBC_BkgTS}, we present as an example, the BGTS distribution we get from one of the sub-threshold CBC candidates (MBTA-1260022599) detected by the LVK collaboration on December 10, 2019, during O3b. The corresponding GW skymap is also shown. One can see that the BGTS distribution contains some dominant spatial features which it inherits from $\omega_\mathrm{k}$. Evident from Fig. \ref{CBC_BkgTS} and \ref{cWB_BkgTS}, it contains several negative TS values as the background neutrinos fall on the pixels belonging to the less-probable regions on the GW skymap. There, $\omega_\mathrm{k}$ is much less than one, resulting in 2~ln($\omega_\mathrm{k}$) to be significantly negative. It acts as a dominant spatial penalty term on the final TS value, obtained from Eq. \ref{TS}.


\subsection{Signal injection and sensitivity studies}
\begin{figure}[!t]
    \centering
    \begin{minipage}{0.5\textwidth}
        \centering
        \includegraphics[width=\textwidth]{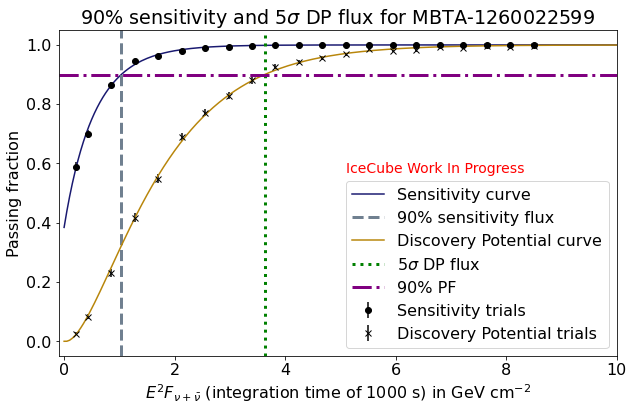}
        \end{minipage}%
    \begin{minipage}{0.5\textwidth}
        \centering
        \includegraphics[width=\textwidth]{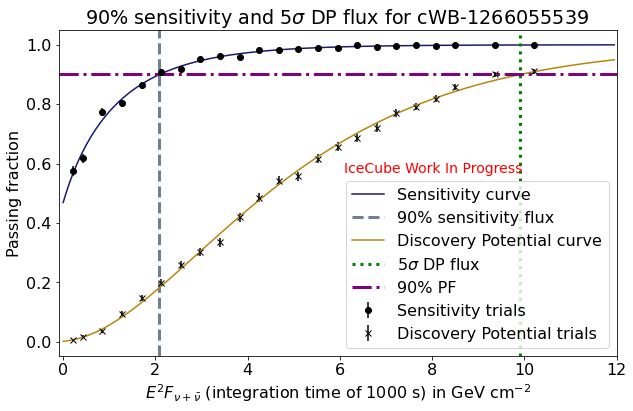}
    \end{minipage}%
	\caption{The 5 $\sigma$ discovery potential flux in comparison to the 90\% sensitivity flux (neutrino + anti-neutrino) for (left) MBTA-1260022599 and (right) cWB-1266055539 respectively. For signal injection, we assume a $E^{-2}$ spectra with spectral index, $\gamma$ = 2. The time-integrated flux, F is computed in a 1000 s time window.}
        \label{sens_DP}
\end{figure}

To perform sensitivity studies, we inject an increasing number of $n_s$, following a spectrum with $\gamma$ = 2, into the GW skymap.
The injected $n_s$ values are scaled to account for the detector's acceptance, allowing us to relate each $n_s$ to a specific flux level. This procedure is repeated across multiple trials for each sub-threshold GW candidate to calculate the Passing Fraction (PF) for various flux levels. The 90\% sensitivity is defined as the flux at which the TS exceeds the median of the BGTS distribution in 90\% of the trials, corresponding to PF~=~0.9. Similarly, 5$\sigma$ discovery potential (DP) is defined as the flux level at which the TS exceeds the threshold corresponding to a 5$\sigma$ (single-sided) fluctuation above the background in 90\% of trials (PF~=~0.9).
In Fig. \ref{sens_DP}, we show the 90\% sensitivity and 5$\sigma$ DP for MBTA-1260022599 and cWB-1266055539 as examples.

\begin{figure}[!t]
\begin{minipage}{0.70\linewidth}
        \includegraphics[width=\linewidth]{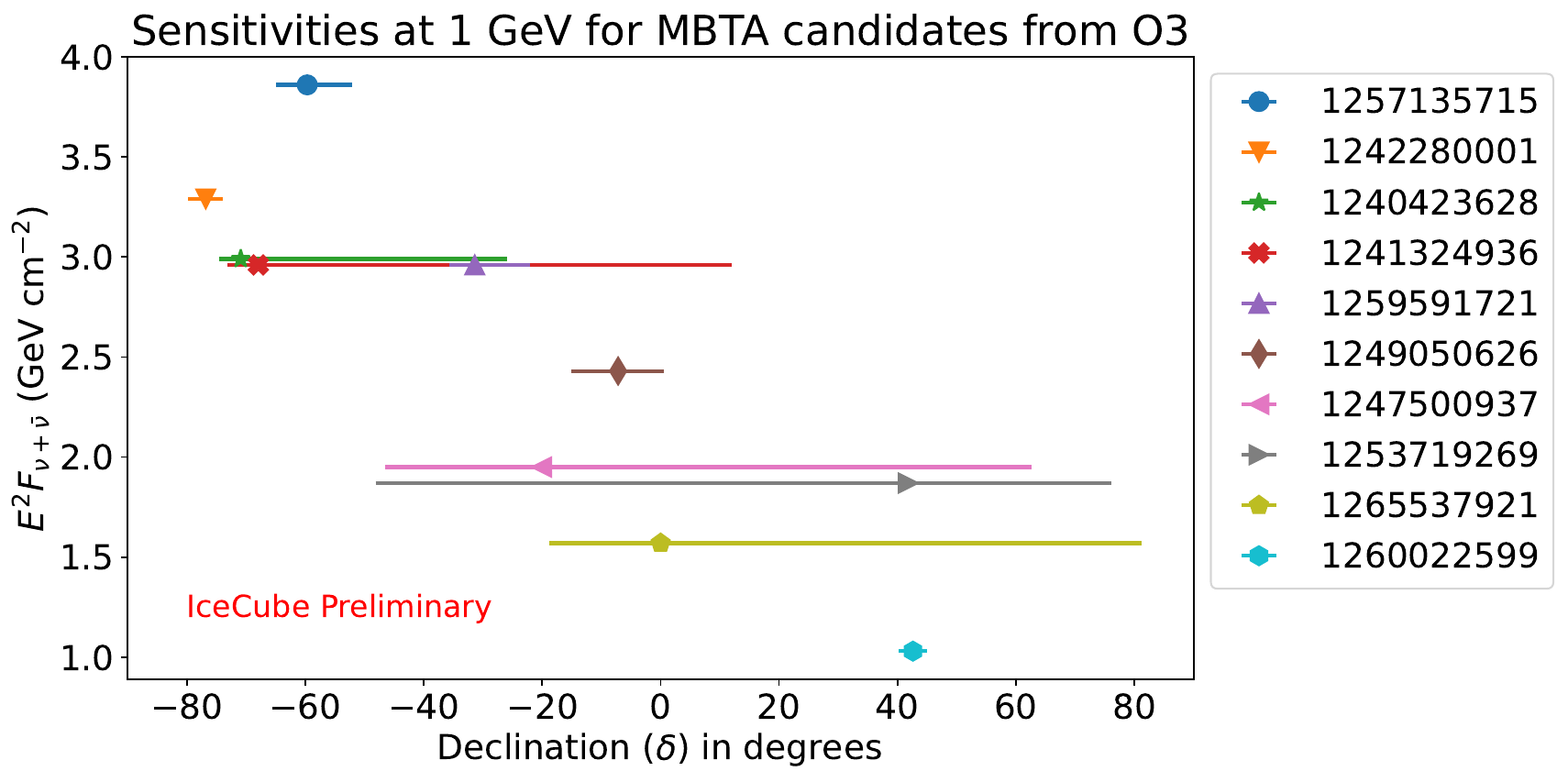}
        \includegraphics[width=\linewidth]{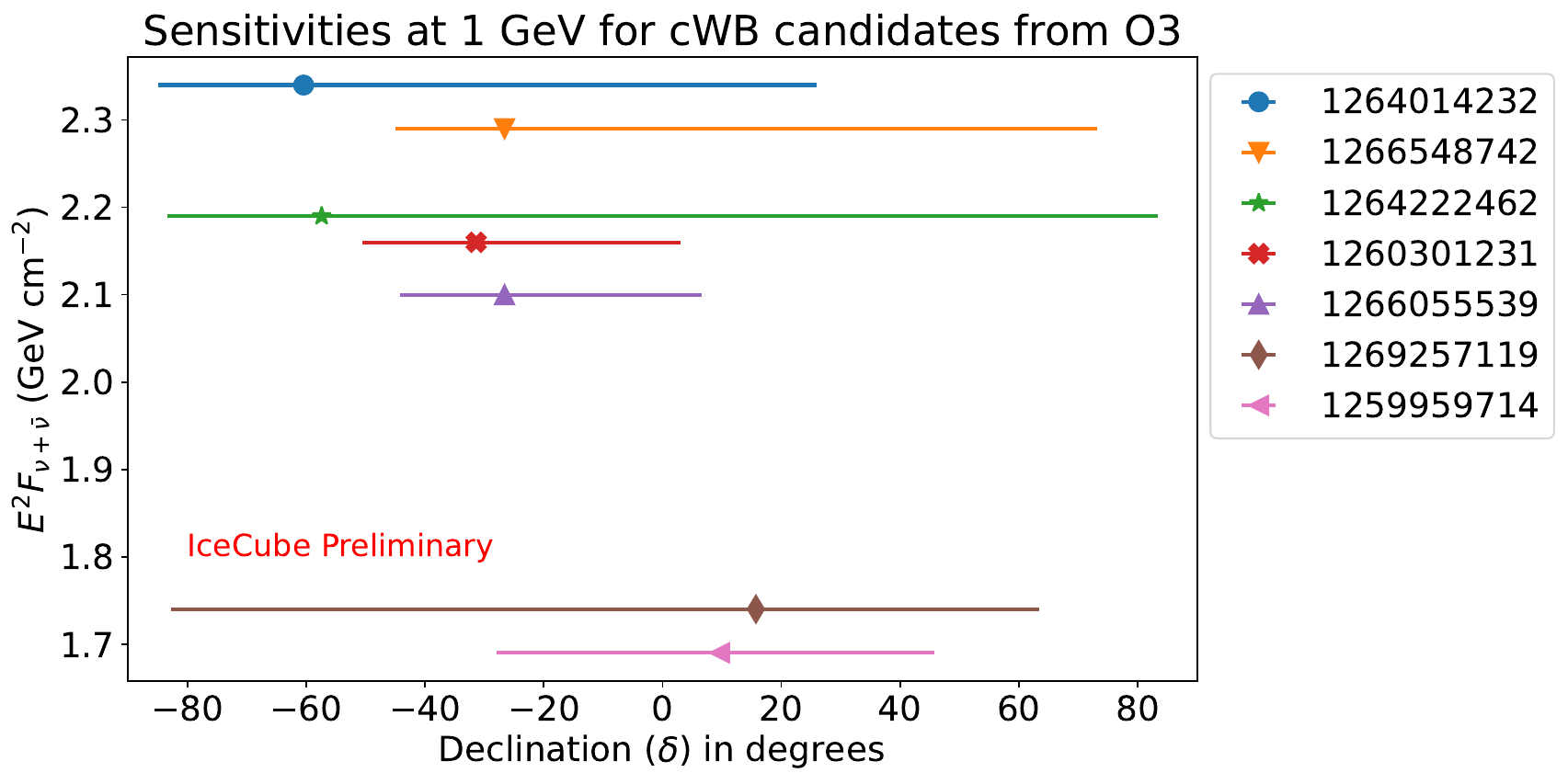}
\end{minipage}
\begin{minipage}{0.29\linewidth}
	\caption{Declination dependence of per-flavour sensitivity flux calculated at 1 GeV for the sub-threshold CBC and cWB candidates identified by (top) MBTA and (bottom) cWB respectively, during O3. The declination of the pixel containing the best-fit source location has been identified by a marker. Different GW candidates have been identified by different coloured markers. The horizontal spread across the markers covers the declination range for 50\% containment area from the respective GW skymap.}
        \label{sensDec}
\end{minipage}
\end{figure}

We also study the declination dependence of per-flavour sensitivity of the sub-threshold GW candidates. We show the sensitivities of the sub-threshold candidates identified by the MBTA and cWB-all sky pipelines in Fig. \ref{sensDec}. This figure demonstrates better sensitivities in the Northern Hemisphere than in the Southern Hemisphere related to atmospheric background. 
Also, it is notable that even the LVK 50\% containment region for most of the sub-threshold candidates is pretty large, ranging from the Northern to the Southern hemisphere. The main reason for this is a worse localisation of sub-threshold candidates compared to that of confident events, by the GW detectors. However, this could be significantly improved if we can identify neutrino counterparts spatially and temporally correlated to these candidates.

\section{Discussion and outlook}
Neutrino counterpart search for confident GW events has been conducted by IceCube over a broad energy range, from a few GeVs to O(100) TeV. In this work, we look for spatially coincident sub-TeV neutrinos with sub-threshold GW candidates, within a 1000 s time window. We have presented per-flavour sensitivity studies for the sub-threshold GW candidates with $p_\mathrm{astro}$ > 0.1. The declination dependence of the per-flavour sensitivity has been determined for candidates from different pipelines. Notably, for some sub-threshold GW candidates, the localisation is particularly poor. Identifying coincident neutrino events would allow us to significantly improve the GW source localisation. This would make future follow-ups of sub-threshold GW candidates more efficient. After the conclusion of our sensitivity study, we will soon unblind the GRECO data sample to search for sub-TeV neutrino emission correlated with archival sub-threshold GW candidates.


\bibliographystyle{apalike}
\bibliography{ref}

\end{document}